\documentstyle[preprint,aps]{revtex}

\begin{document}

\draft

\title{ Comment on  \\ 
``Measurement of the $\pi^+ \vec p$ analyzing power at 68.3 MeV"}
\author{Richard A. Arndt, Igor I. Strakovsky$^\dagger$ and Ron L. Workman}
\address{Department of Physics, Virginia Polytechnic Institute and State
University, Blacksburg, VA 24061}
 
\date{\today}
\maketitle

\begin{abstract}

We comment on a recent paper by Weiser et al.~[Phys. Rev. C {\bf 54}, 
1930 (1996)]. The authors have performed a single-energy analysis of
$\pi^+ p$ scattering data at 68.3 MeV, finding a value for the $S_{31}$
phase shift about 1$^{\rm o}$ smaller than found in the Karlsruhe-Helsinki 
(KH) partial-wave analysis. The authors use this result to argue for
a dispersion relation analysis using recently measured data, so that
their effect on the $\pi NN$ coupling constant ($f^2$) and $\Sigma$ 
amplitude can be determined. We note that these tasks were accomplished 
prior to the submission of the above paper. We clarify the effect of this 
new analyzing power data on $f^2$ and the $\Sigma$ amplitude.  

\end{abstract}

\vskip .5cm


\narrowtext

The authors of Ref.\cite{Weiser} have used their new measurement of 
the $\pi^+ \vec p$ analyzing power at 68.3 MeV in a determination of 
the S31 phase shift. The result is a value about one degree smaller 
than that found in the KH analysis\cite{KH80}. Given this discrepancy,
the  authors suggest the need for an alternative dispersion analysis of
$\pi N$ scattering data to determine the $\Sigma$ amplitude and $f^2$. 

The $\Sigma$ amplitude can be determined by extrapolating the $A^{(+)}$ 
amplitude\cite{KH80} to the Cheng-Dashen point ($\nu = 0$, $t= 2\mu^2$). 
A reliable extrapolation requires a precise determination of $A^{(+)}$ at
low energies. This motivated the low-energy measurements of 
Ref.\cite{Weiser}. The effect of low-energy data on $f^2$ can be seen from
the Goldberger-Miyazawa-Oehme (GMO) sum rule\cite{GMO}, 
which relates the S-wave scattering lengths to $f^2$ and an integral over 
total cross sections. 

In this Comment, we describe the   
effect of the new analyzing power data\cite{Weiser}, on the  
$\Sigma$ amplitude and $f^2$. These questions have already been considered 
by Arndt {\it et al.}\cite{SM95,Arndt} and indirectly by Sainio\cite{Sainio}.  
In the analyses of Arndt\cite{Arndt}, the result  
$\Sigma \approx 68$ MeV was found. Sainio\cite{Sainio}, using the solutions 
KH80\cite{KH80} and SM95\cite{SM95}, 
gave a range ($\Sigma$ = $60\pm 10$ MeV) consistent with our estimate. 
(The SM95 analysis and its associated single-energy analyses
utilized the data of Ref.\cite{Weiser}.) 
Values for $f^2$ were also determined\cite{Arndt} using solution 
SM95. The range of values ($f^2 = 0.076\pm 0.001$) was found to 
lie within the range ($f^2 = 0.076\pm 0.003$) quoted in Ref.\cite{Weiser}. 

In Table~I, we compare the phase shifts and $\chi^2$ values from a number
of analyses\cite{KH80,SM95,C6,SM90,KA84}. The renormalization factor is
also given for each of the two data sets (Set I containing 3 points and
Set II containing 4). (Note that this is the factor which should be applied
to the analysis when it is compared to the data.) In all cases, the  
analyses are renormalized downward to fit the data. Again, in all cases,
the data of Set II are in better agreement with the analyses. 

As should be expected, the single-energy 
analysis\cite{C6} (C6) has the lowest $\chi^2$. Here the renormalization 
factor is near unity. In fact, the phase shifts found in Ref.\cite{Weiser}
almost perfectly reproduce our results. However, when the data of 
Ref.\cite{Weiser} were included in an energy-dependent analysis, 
more renormalization was necessary. This in turn resulted in an S31 phase 
which was closer to the KH result. The S31 phase found in solution 
SM90\cite{SM90} (prior to the measurement of Ref.\cite{Weiser}
and prior to the addition of dispersion relation constraints) is not very
different from SM95\cite{SM95} 
(which included the preliminary data of Ref.\cite{Weiser}
and applied a range of dispersion relation constraints). 

Finally, in Table~II of Ref.\cite{Weiser}, the authors claim that the KH 
analysis gives a $\chi^2$ of 116.54 against the 7 analyzing power data. 
This very high $\chi^2$ results if the systematic error ($\pm$5\%) is 
neglected. When the systematic error is taken into account, the $\chi^2$ 
drops by more than a factor of 5.

\acknowledgments 

We thank R. Weiser for providing data prior to publication and for
communications regarding the systematic uncertainties of the
experiment. I.S. acknowledges the hospitality extended by the Physics 
Department of Virginia Tech.  
This work was supported in part by the U.S.~Department of Energy Grant 
DE--FG05--88ER40454. 

%%%%%%%%%%%%%%%%%%%%%%%%%%%%%%%%%%%%%%%%%%%%%%%%%%%%%%%%%%%%%%%%%
%%%                      References
%%%%%%%%%%%%%%%%%%%%%%%%%%%%%%%%%%%%%%%%%%%%%%%%%%%%%%%%%%%%%%%%%%

%%%%%%%%%%%%%%%%%%%%%%%%%%%%%%%%%%%%%%%%%%%%%%%%%%%%%%%%%%%%%%%%%
%%%                      Tables
%%%%%%%%%%%%%%%%%%%%%%%%%%%%%%%%%%%%%%%%%%%%%%%%%%%%%%%%%%%%%%%%%%
\newpage
\mediumtext
\vfill
\eject
Table~I. Fits of partial-wave analyses to the analyzing power data of
Ref.\cite{Weiser}. Results for the S31, P31, and P33 partial-waves are
given. Types of analysis are: DR (utilizing dispersion relation
constraints), ED (energy-dependent with no dispersion relation constraints), 
and SE (single-energy). A predicted renormalization factor (see text) is also 
given in each case for the two separate data sets.

\vskip 10pt
\centerline{
\vbox{\offinterlineskip
\hrule
\hrule
\halign{\hfill#\hfill&\qquad\hfill#\hfill&\qquad\hfill#\hfill
&\qquad\hfill#\hfill&\qquad\hfill#\hfill&\qquad\hfill#\hfill
&\qquad\hfill#\hfill\cr
\noalign{\vskip 6pt} %
%The title line
PWA&Type&S31&P31&P33&$\chi^2$/datum&(Norm I, Norm II) \cr
\noalign{\vskip 6pt}
\noalign{\hrule}
\noalign{\vskip 10pt}
KH80\cite{KH80} & DR &  -6.96  & -1.23 & 10.12 & 21.3/7 & (0.86, 0.87) \cr
\noalign{\vskip 6pt} 
KA84\cite{KA84} & DR &  -6.96  & -1.27 & 10.08 & 21.8/7 & (0.85, 0.87) \cr
\noalign{\vskip 6pt}
SM90\cite{SM90} & ED &  -6.54  & -1.12 & 9.99  & 10.1/7 & (0.91, 0.93) \cr
\noalign{\vskip 6pt}
SM95\cite{SM95} & DR &  -6.43  & -1.23 & 9.94  & 8.3/7 & (0.92, 0.94) \cr
\noalign{\vskip 6pt}
C6\cite{C6}  & SE    &  -6.08  & -1.23 & 9.65  & 4.3/7 & (0.96, 0.99) \cr   
\noalign{\vskip 10pt}}
\hrule}}
\vfill
\eject

\begin{references}
\bibitem{C6} 
This single-energy solution was described in 
Ref.\cite{SM95} and was fit to data between 61 MeV and 69 MeV. 
Preliminary data from Ref.\cite{Weiser} were included.
\end{references}
\end{document}